# NEO Characterization Science Case
# For a
# Low Resolution Spectrograph

Mark Trueblood
Winer Observatory (648)
P.O. Box 797, Sonoita, AZ  85637-0797 USA
winer.obs@gmail.com

Larry Lebofsky
Planetary Science Institute
1700 E. Ft. Lowell, Suite 106, Tucson, AZ  85719-2395 USA
lebofsky@psi.edu

Robert Crawford
Rincon Ranch Observatory
2853 S. Quail Trail, Tucson, AZ  85730-5627 USA
rincon-ranch@earthlink.net

**Abstract:** Near Earth Asteroids (NEAs) and dead comets comprise the vast majority of the population of Near Earth Objects (NEOs) detected to date. Less is known of their physical properties than of the much larger population of main-belt asteroids. Due to the faintness and short duration of visibility of NEOs, many characterization studies use broadband filters in 3 to 8 colors for taxonomic classification and to study surface chemical composition. A spectrograph with low spectral resolution R~30 used in a campaign or a continuing program on a small telescope (1-2m class) would vastly improve the quantity and quality of data on NEOs. The proposed baseline instrument would work in the visible using a CCD detector, with a possible upgrade to include a second, near-IR (NIR) channel extending coverage to 2.5 µm or beyond. The optical design needs to optimize overall optical throughput to permit observation of the faintest possible objects on small telescopes at acceptable signal-to-noise (S/N) ratios. An imaging mode to obtain accurate same-night broadband photometry would be highly desirable.

## 1. Issues and Questions in NEO Characterization

Asteroid studies help reveal the processes and conditions of solar, stellar, and planet formation, including accretion of dust grains into planetesimals by collisions in the primordial nebula. One can begin to understand these early processes and conditions through determination of asteroid mineral composition and knowledge of the formation temperatures of the minerals. Earth has hidden the physical evidence of these processes by differentiation, erosion, metamorphism, and remelting, so one must turn to meteorites and asteroids to understand the solar system formation processes (Gaffey *et al.* 1993). Although great progress has been made in recent years in understanding asteroids and comets in general, using both ground-based facilities and space probes, the sub-class of these bodies that wander near the Earth, the NEOs, is less well studied, and therefore less well understood.

NEOs are asteroids or comets with perihelion distances $q<1.3$ AU. The vast majority are NEAs and dead comets (with no detectable surface ices) masquerading as asteroids (Remo 1994), with the remainder being active comets. NEOs are not as well characterized as main belt asteroids because they tend to be smaller and fainter, and they tend to fade rapidly (a few days to weeks) in brightness after discovery as a function of the NEO-Earth-Sun geometry.

As with the main-belt asteroids, there are several ways to classify NEOs, including by orbital elements, brightness or size, rotation period, color, and chemical composition. All of these have been used in practice (e.g., Hirayama 1922, Chapman *et al.* 1994, and Binzel 2002). A primary purpose for classification by orbital parameters is to determine if the body poses a threat for collision with Earth (e.g., NASA/JPL PHA Web site, http://neo.jpl.nasa.gov/neo/pha.html). Orbital parameters can also be used to infer how the NEOs originate if the mechanism is the collision of two main-belt objects (Bottke *et al.* 2002). Classification by brightness or size is summarized in the object's H magnitude, which is its brightness at a standardized distance and phase angle. Each H magnitude corresponds to a factor-of-two range in diameter depending on the albedo of the object, which is seldom known with accuracy (Binzel 2010).

Broadband (or even unfiltered) photometry in one color can yield a rotation period, which can indicate structure and other physical properties. Figure 1 below shows a rotation light curve for the NEO 25143 Itokawa, obtained by Brian Warner of the Palmer Divide Observatory, that gives a rotation period of $12.09 \pm 0.01$ hours. For an object with diameter >150 meters, a rotation period faster than ~2 hours means that the object is most likely not a rubble pile, but must be bound by internal forces stronger than gravity (Pravec *et al.* 2002, Pravec and Harris 2000, Harris 1996). This type of information is interesting not only to those piecing together the larger picture of the NEO population, but also to those who may someday plan an intercept mission to deflect or destroy a hazardous NEO, since the energy coupling constant depends heavily on the type of material composing the body (Shafer *et al.* 1994).

Color filter photometry can be used to obtain information on chemical composition and to aid in classifying the body according to one of several taxonomic systems. Very early surveys began with photographic plates, but quickly changed to more sensitive photomultiplier tubes when they became available (Groeneveld and Kuiper 1954). Early observations using the McCord double-beam photometer employed a photomultiplier tube as the detector with 25 filters (to define a 24-color system) to classify nearly 300 main-belt asteroids (Chapman and Gaffey 1979). Although filter photometry has been used reliably and successfully to classify asteroids (Tedesco *et al.* 1982), it works best on the slower-rotating bodies, when there is sufficient time to image the same face of the rotating body in all the colors.

One very widely-used asteroid classification system (Tholen 1984) was defined using 8 filters and photomultiplier tubes with sensitivities out to the NIR, beyond the current capabilities of CCDs. The Tholen "Eight Color Asteroid Classification" (ECAC) system was used with success on hundreds of main belt minor planets (Zellner *et al.* 1985). The eight colors of the system classify asteroids into 3 groups and up to 14 classes, although the three classes of the X Group have similar, generally featureless spectra and require independent measurement of the albedo to complete the classification. Modern analogs using CCDs and NIR detectors are now in use, but tend to use broadband filters suitable to the detector involved (e.g., Hicks et al. 2009B).

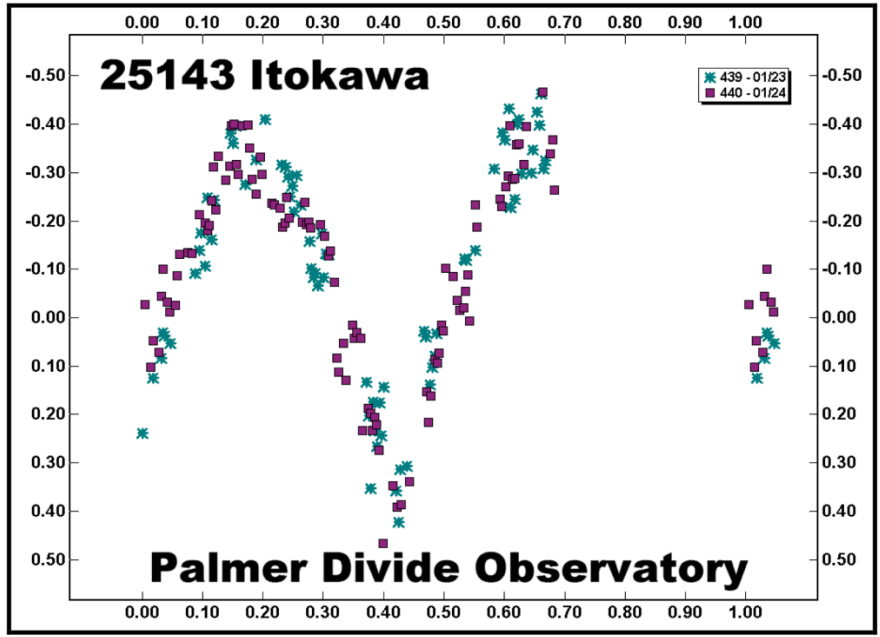

| Asteroid: | 25143 Itokawa | First Session: | 2004 01 23 |
|---|---|---|---|
| Period: | 12.09h ±0.01h | Last Session: | 2004 01 24 |
| Amplitude: | 0.70m ±0.03m | Total Sessions: | 2 (178 data points) |
| | | Equipment: | 0.5m f/8.1 R/C FLI-1001E |
| | | Temp/Exposure: | -30°C / 90s |

| Discovered: | September 26, 1998 LINEAR at Socorro |
|---|---|
| Class: | |
| H: | 19.20 |
| SM Axis: | 1.323 |
| Inclination: | 1.728 |
| Eccentricity: | 0.2795 |

**NOTES:**
Naming information not available in Schamdel, 5$^{th}$ ed.

This asteroid will be studied in 2004 for evidence of the YORP effect. See "Detectability of YORP rotational slowing of asteroid 25143 Itokawa" by D. Korkrouhlicky et al in Astronomy and Astrophysics. The data obtained by Palmer Divide *seems* to confirm the theory that the asteroid's rotation will have slowed enough to cause an approximately one hour phase shift since 2001. However, more detailed observations must be acquired to confirm this since the asteroid is just outside the "tumbler zone" as defined by Pravec and Harris. If tumbling, even slightly, the observed delay could be the result of that and not YORP.

**Figure 1. Rotation light curve obtained from photometry of 25143 Itokawa. From Warner 2004.**

A system with increased spectral resolution was used in the 52-color NIR survey (Bell *et al.* 1988). In this survey, a circular-variable-bandpass interference filter was rotated to obtain 52 bandpasses between 0.8 µm and 2.5 µm. This was used on the 3-m IRTF between 1983 and 1987 to obtain 143 full 52-channel spectra (photometry) of 119 different asteroids. These data were used to infer mineralogy of the bodies and to conclude that S-type asteroids varied widely in their olivine/pyroxene ratios. Possible sources of error with this system include variations in color across a non-zero diameter beam, and tolerances in returning to precisely the same wavelength position on the continuously-varying filter.

A limitation of systems based on multiple color filters is that each filter discards photons outside its bandpass, which can be narrow if eight or more colors are placed across the CCD response band. There are two issues raised by this. The first is that the NEO is rotating while one is observing in one band, so that the target body presents a different "face," with a potentially different mineralogical composition and albedo, when the last filter is reached. This is further complicated by the need for separate calibration frames for each different filter. The second is that when conducting a survey or campaign on a large number of objects, efficiency becomes important, and discarding out-of-band flux is an inefficient use of perfectly good photons. On the positive side, color filter photometry can yield very high S/N ratios, and broadband filters can reach fainter objects than even low spectral resolution spectrographs.

A more efficient means of obtaining the same information is to use a very low resolution spectrograph. Such an instrument with a spectral resolution R ~10 could theoretically duplicate the Tholen ECAC filter system, while a slightly higher R of 20 to 30 would yield more information without a large decrease in sensitivity (limiting magnitude). The requirements for success in spectroscopic observation are to be able to observe faint bodies in the visible (using a CCD detector) on a small telescope before the body rotates too much, while achieving a reasonable S/N ratio for spectroscopy. We define the terms "small telescope," "rotates too much," and "reasonable S/N ratio," as well as consider relevant design trades, in Section 3 below.

With such an instrument, one is not trying to detect detailed spectral features. Rather, the interest is in the overall shape and slope of the spectral curve and, if resolution permits, the presence of major spectral features indicative of composition. Given even crude spectral resolution, one can distinguish one taxonomic class of asteroid from another with sufficient clarity to make reliable classifications using the Tholen ECAC system (and with the same limitation that albedo must be independently determined to remove the degeneracy that otherwise exists among X-Group objects). One must be able to calibrate the spectrum to within at least 5% to obtain adequate spectro-photometric accuracy.

Figure 2 (from Bus *et al.* 2002) compares spectra obtained in the Small Main-belt Asteroid Spectroscopic Survey (SMASS) II survey to the photometry and classifications in the Tholen ECAC system. The figure also demonstrates some of the spectral shapes that one needs to distinguish for taxonomic classification. Objects in the leftmost and rightmost columns can be comparably classified using either SMASS II spectroscopy or ECAC photometry. The E, M, and P class objects in the center belong to the Tholen X Group, for which the absence of useful spectral features precludes a meaningful classification without independent knowledge of the albedo. Given this degeneracy, it is necessary to measure the albedo to further understand X Group objects, which display a wide range from dark to bright (Harris and Lagerros 2002).

Traditionally, albedo has been measured in the mid-IR, where most asteroids are up to 10 magnitudes brighter than in the visible. Mid-IR observations are combined with visible observations and thermal models (Lebofsky *et al*. 1986) to determine both the size and albedo simultaneously, based on the contrast of the object's brightness in reflected sunlight and at thermal wavelengths. However, mid-IR instrumentation is expensive and the observations are difficult to make, requiring chopping with the secondary mirror and nodding of the telescope. Rivkin *et al*. (2003) demonstrated how NEO albedos may be *constrained* (but not necessarily determined) using NIR spectra taken in the region 1–2.5 µm, the phase angle, and solar distance.

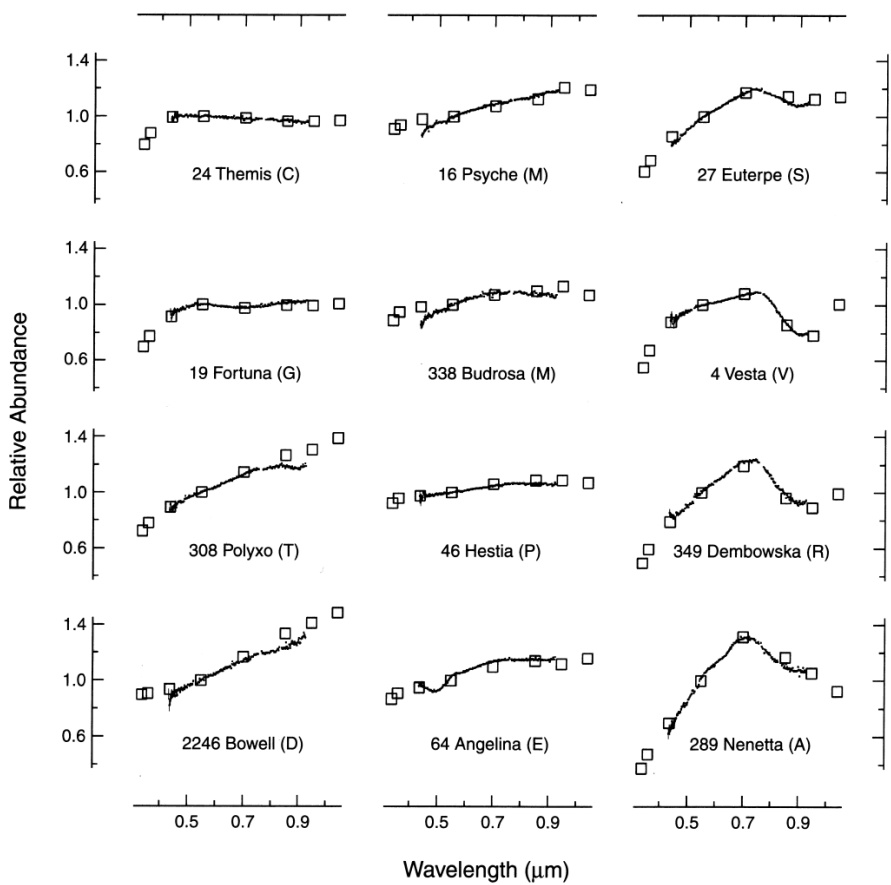

**Fig. 2.** A comparison between the Eight-Color Asteroid Survey (ECAS) colors (open squares) and SMASSII spectra (closed dots with error bars) for selected asteroids in various Tholen taxonomic classes. Those asteroids along the middle column (16 Psyche, 338 Budrosa, 46 Hestia, and 64 Angelina) show the range of spectral diversity among the Tholen X-class (E, M, and P classes, which can only be separated based on albedo) asteroids observed in the SMASSII data. Minor absorption features that are visible include the broad 0.7-µm band in the spectrum of 19 Fortuna, a 0.49-µm band in the spectrum of 64 Angelina, and a subtle 0.65-µm feature in the A-type spectrum of 289 Nenetta.

Apart from inquiry-driven science, there are practical reasons for studying NEOs. If an NEO is discovered to be a threat to Earth, knowledge of its physical properties is essential before a mitigation strategy can be selected (Shafer 1994). For example, how it will react to irradiation from an Earth-based or space-based laser will depend on its mass and its composition, that is, how its material absorbs or reflects the laser light. The same can be said of other mitigation techniques that have been proposed, such as thermonuclear explosives detonated at some distance from the surface of the body (where X-rays and neutrons heat the surface and ablate material that then provides an impulse to the body), or even exploding conventional high explosives or a thermonuclear device below the surface. The energy coupling constants for different types of explosives delivered by various means depend on the material composition of the asteroid (Shafer 1994).

The discussion in this section demonstrates that both planetary and stellar system formation science and the practical aspects of NEO hazard mitigation require the characterization of NEOs. Spectro-photometry reveals the surface chemical composition of NEOs, which enables taxonomy, which then assists studies of collision histories and of the origins of NEOs with similar taxonomic classifications. Furthermore, knowledge of NEO composition reveals important clues to the conditions under which the components of the solar system were formed. Thus, spectro-photometric study of NEOs is an important means of understanding the origins and nature of NEOs and other solar system bodies.

## 2. Observing Methods and Available Data

The most commonly available NEO data are astrometry (used to compute orbital elements) and absolute brightness (H), which are obtained through the NEO surveys and those involved in astrometric follow-up observations. Astronomers have obtained light curves of many asteroids, but these tend to be the larger (and brighter) main belt asteroids. Only a few observers obtain rotation light curves, filter photometry, or spectra of NEOs on a regular and ongoing basis. This is due in part to the relative faintness of NEOs (typically $18<V<23$) and the fact that they fade rapidly after discovery.

One result of this situation is that H values for NEOs tend to be in error by 0.5 magnitudes or more. The authors observe NEOs for astrometric follow-up of discoveries and are familiar with the performance of software for astrometric data reduction. Such software reports astrometry with sub-arc second accuracy, but reported photometric values can vary by 0.5 magnitudes between images of the same object taken only a few minutes apart. Binzel (2010) has long advocated a program of accurate, well calibrated photometry of NEOs to obtain accurate H magnitudes. These would be useful for population statistics, impact rates, and assessment of the overall Earth hazard (Binzel 2010).

Chapman *et al.* (1994) present a table of NEA properties for 84 objects. This table was updated in Binzel (2002) and now lists 791 objects (http://earn.dir.de/nea/table1_new.html) as of July 2010. Note that in 1994, only 56 of the 84 objects (67%) were classified by mineralogical composition, and two of the classifications were questionable. In the July 2010 web-based table, only 396 (50%) of the total 791 objects had taxonomic classifications, even though 419 objects (53%) had been observed enough to become numbered asteroids. Although the total number of objects has grown substantially (from 84 to 791) in the last 16 years, the fraction with taxonomy has not kept pace. This is a result of many factors, including the decreasing brightness of new discoveries, the difficulty of obtaining the telescope time required to perform characterization studies, and the lack of proper instrumentation needed to obtain adequate data for taxonomic classification.

Filter photometry using 3 to 8 broadband filters is still the most frequently used method of characterizing NEOs. The largest database of asteroid filter photometry is the Sloan Digital Sky Survey (SDSS) Moving Object Catalog (MOC) containing 471,569 moving objects observed prior to March 2007 (http://www.astro.washington.edu/users/ivezic/sdssmoc/sdssmoc.html). Of those, 220,101 are linked to 104,449 unique objects in the Lowell Observatory ASTORB file, for which orbital elements are also listed. All SDSS objects were observed in the five SDSS filter colors defined for the survey (Fukugita *et al.* 1996). Although this extremely large database gives a glimpse into the colors of hundreds of thousands of asteroids, with only five colors in the survey, the effective spectral resolution (R ~5) is too low for complete and reliable taxonomic classification using any of the standard systems now in common use, although one may place most objects into more basic taxonomic classes (Bus 2010).

The SDSS MOC was a serendipitous consequence of the SDSS survey, which was directed toward other objectives. An example of the type of study one can do with these data is provided by Moskovitz *et al.* 2009, in which they find candidate members of the Vesta (V-type) family using the SDSS MOC V1.0, then follow up those candidates using spectroscopy on the Keck II and University of Hawai'i 2.2-meter telescopes.

Some recent studies using filter photometry of asteroids are focused on specific targets of scientific interest. For example, Hicks *et al.* 2009A obtained Bessel BVRI photometry of potentially hazardous asteroid (PHA) 2009 KC3 using the 0.6-m telescope at Table Mountain Observatory and compared their resulting colors with those from the spectroscopic SMASS II survey (Bus and Binzel 2002). Weaver *et al.* 2009 used a combination of Hubble Space Telescope filter photometry in the range ~150 nm to ~700 nm and ground-based spectroscopy to prepare for a flyby of the asteroid Lutetia by the Rosetta spacecraft. Masiero *et al.* 2009 used the 3.6-meter Canada-France-Hawaii Telescope on Mauna Kea and the MegaPrime/MegaCam large format camera combination to obtain very wide field images through single SDSS filters (either $g'$ or $r'$) to conduct an unbiased survey to detect asteroids. The untargeted survey found 828 Main Belt asteroids to a limiting magnitude of $g'$ ~22.5 in a corresponding diameter range of 0.4<D<10 km. They were able to perform rotation period fits to 278 of these objects, which was the goal of the survey.

A study using only two filters minimizes the problem of observing rapidly-rotating bodies and, for faint objects, an imaging study using broadband filters could go fainter than a spectroscopic study (Bus 2010). A campaign or survey using up to five broadband filters would be effective in studying all NEOs, even the rapid rotators, could reach fainter limiting magnitudes than a low-R spectrograph, and would be easier to implement on a robotic telescope. However, the broadband filter approach yields less total information about the spectrum than a low-R spectrograph, so resolving questionable cases of taxonomic classification would require follow-up using a low-R spectrograph. Simulations would be needed to determine which approach is more efficient in terms of total telescope time to observe a given number of objects.

Gaffey *et al.* (1993) reviewed asteroid reflectance spectroscopy, the technique that yields the most information about NEOs. The problem with spectroscopy at most observatories is that the majority of instruments are built for bright objects or for fainter objects on large telescopes, where it is difficult to obtain time for asteroid characterization studies. These instruments tend to have higher spectral resolution (e.g., R ~1000 or higher) than is required for simple asteroid classification. Such high spectral resolution places a relatively bright limit on the population of asteroids that can be observed. As a result, most asteroid spectroscopy is confined to brighter objects, leaving NEOs largely unobserved using spectroscopic techniques.

Despite these limitations, many modern asteroid researchers have moved from filter photometry to low resolution spectroscopy to obtain data for taxonomic classification surveys. For example, Bus and Binzel 2002A,B conducted the Small Main-belt Asteroid Spectroscopic Survey (SMASS) II. They obtained low spectral resolution (R ~100) CCD spectra for 1447 main-belt and near-Earth asteroids over the spectral region 0.4 to 0.9 μm and used them to define 26 taxonomic classes. They employed the Mark III long slit grism spectrograph mounted on the 2.4-meter Hiltner and the 1.3-meter McGraw-Hill telescopes at the MDM Observatory on Kitt Peak. The faintest asteroid in the published data was approximately V ~18.5. Although the SMASS II survey contributed significantly to our understanding of both main belt asteroids and NEOs, the relatively small number of NEOs included in the survey (106) is due to their relative faintness compared to the more accessible population of main belt objects.

Having obtained a good sample of main-belt objects in earlier observations, the latter part of the SMASS II program focused on NEOs and Mars crossers (MCs). Binzel *et al.* 2004 contains visible and NIR spectra on 252 NEOs and MCs, classification of the observed objects using the Bus (1999, Ph.D. thesis) 26-type taxonomic system, and the characteristics of these objects. Several hundred NEO spectra in the range 0.8 to 2.5 μm were obtained using SpeX on the IRTF, and the spectra that have been released (60%) are available at http://smass.mit.edu/minus.html (Binzel 2010).

Despite the efforts of Binzel, Bus, and their collaborators, as well as continuing efforts by Licandro *et al.* 2008 and others, there remains a shortage of data for NEO characterization, particularly for taxonomy, compared to the main asteroid belt. Astrometric follow-up of NEO discovery surveys is done routinely on telescopes with apertures in the range 0.3-meter to 2.4 meters, as the Minor Planet Electronic Circulars (MPECs) issued daily by the Minor Planet Center demonstrate (http://www.minorplanetcenter.org/mpec/RecentMPECs.html). As the brighter NEOs are discovered, follow-up by amateurs with smaller telescopes becomes increasingly more difficult, and even at present some of the fainter objects are not being adequately followed up.

An examination of current MPECs indicates that NEOs now being discovered range in brightness approximately 18<V<23, and there are many NEOs on the JPL web site (http://neo.jpl.nasa.gov/neo/) in need of taxonomic classification in the range 16<V<18. For example, the JPL web site (August 2010) showed 2938 NEOs with H <=20, 1018 NEOs with H <=18, and 185 NEOs with H <=16. These facts indicate that a low-R spectrograph mounted on a small telescope would be useful in obtaining data on a large and interesting fraction of the NEO population.

The SMASS II survey was highly successful in gathering the data needed to classify a large number of main belt asteroids and many NEOs and MCs, and to provide data on a sufficiently large number of bodies to give a statistically significant sample, at least of the larger and brighter objects. Despite the fact that an imager with a few broadband filters can provide a basic taxonomic capability capable of reaching fainter limiting magnitudes, we contend that a modern, efficient low-R spectrograph for the same class of small telescopes used in the SMASS II survey is needed to extend prior studies to the entire NEO population in a manner that provides a complete and unambiguous dataset. It is these less-heavily subscribed telescopes that are more likely to be available to perform a similar type of survey for NEOs.

In Section 1, we showed how spectro-photometry was key to understanding the origin and nature of NEOs. In this section, we have argued that although sufficient information to categorize NEOs may be obtained using filter photometry, a low resolution spectrograph would provide the most efficient data gathering method for an extended survey of NEO composition on small telescopes. Such a survey, if properly calibrated, would provide a robust dataset useful for both taxonomic purposes, as well as other scientific and practical applications.

## 3. Desired Instrumentation for NEO Taxonomic Studies

Given that existing programs are providing astrometric and photometric data to determine orbital parameters and rotation rates, a unique and important contribution to solar system studies can be made by a low-R spectrograph designed for NEO taxonomic studies on small (e.g., 1-2m class) telescopes. Such an instrument would have the following characteristics, which are discussed in greater detail below:

- One channel using a CCD detector able to capture the entire visible light band (0.36 – 1.0 μm wavelength regime) in a single exposure.
- Extremely high optical throughput, exceeding 70% (peak) for the instrument itself (not counting losses in the atmosphere, telescope, or detector).
- Spectral resolution R in the range 10<R<100, with R ~30 thought to be most desirable to support taxonomic studies.
- Approximately constant R with wavelength (variation of less than 30% over the wavelength range).
- Ability to obtain an S/N ratio of 20-30 on a 1.5-m telescope in 20 minutes on a V ~20 object.
- Built-in spectral and photometric calibration standards.
- Fewest possible moving parts for simplicity and stability.
- Ability to guide to keep non-sidereal targets on the slit.
- Operated remotely by software at the observer's console in the control room.
- Capable of robotic operation, if possible.
- Optionally, a direct imaging mode to enable "same-night" precision photometry of the targets whose spectra are recorded and of standard stars for calibration.

- Data reduction pipeline available from the moment the instrument arrives at the telescope for commissioning.

It is important to state that the discussions below assume that a proper design study will be performed before the instrument we conceptualize is designed in detail to meet these requirements. Such a design study will weigh the demands of NEO taxonomy against the realities of available technologies to determine what is feasible within realistic budget constraints. The Principal Investigator or Project Scientist would then determine whether the resulting instrument is worth building before proceeding with the project.

Although considerable work has been done on asteroids in the NIR and mid-IR, taxonomic classification depends primarily on spectra taken in the visible, as was shown in Figure 2, so that a one-channel instrument using a CCD detector would be well suited for the task. The instrument should give good image quality in the region 0.4 – 0.9 μm as a requirement and 0.36 – 1.0 μm as a goal. For example, there are features below 0.4 μm and near 0.95 μm that are interesting and useful for taxonomic classification. Modern anti-reflection coatings on CCDs can be obtained to give good response over the wider spectral range out to the silicon cutoff. Two visible channels with separate detectors (with different response curves, one blue-sensitive and one red-sensitive) could be employed if funding permitted, although this would be a hard-to-justify luxury given the instrument's intended use on small telescopes.

What is crucial is that the entire visible band be obtained in a single integration. The target bodies will rotate with periods as short as two hours, and the spectra must represent a single face on the surface of the body. This requirement points toward a prism spectrograph with fast read-out detector electronics. A grism spectrograph will have overlap in the contiguous grating/grism orders whenever the wavelength range exceeds a factor of 2, which it would in covering even the minimum wavelength range of 0.4 – 0.9 μm. It is not acceptable to take an exposure over half the wavelength range, change an element inside the instrument (e.g., insert an order sorting filter), then take a second exposure to obtain the remaining wavelength range.

Rapidly rotating bodies may require exposures as short as 10-15 minutes, depending on their rotation rates, and short exposures will also limit the number of cosmic ray events on the CCD detector (Bus and Binzel 2002A). A short exposure time has serious implications for the tradeoff of spectral resolution R versus limiting magnitude for a given aperture telescope (discussed below). It also means that both the detector and its controller should be optimized for rapid readout. Modern CCDs, even those with relatively large formats, usually can be read out in 20 seconds or less at spectroscopic S/N ratios, which tend to demand lower read noise than imaging applications.

Given that very faint objects (18<V<23) will be observed on small (1-m to 2.4-m) telescopes, a very efficient instrument is needed with extremely high throughput, exceeding 70% at the peak (instrumental throughput, not counting atmosphere, telescope, or detector efficiencies). This can be achieved using one (or possibly a few) prisms for the dispersing element, but most likely not with gratings, grisms, or a combination of these, which typically have total instrument optical efficiencies at blaze of 40% or less. Even VPH grisms/gratings are unlikely to yield the required throughput efficiency across the entire CCD detection bandpass.

Spectral resolution R can be a difficult trade. One always wants the increased visibility into physical processes of interest that higher R provides, but that visibility comes at a price – a brighter limiting magnitude of objects one can study. This results in fewer possible targets. An R of 10 essentially duplicates what can be done using broadband filters (although it retains the advantage of needing only a single exposure to cover the wavelength range), while an R of 100 begins to limit the number of objects one can study with a small telescope. A problem to be faced during instrument design is that it is difficult to calibrate the wavelength and to correct properly for telluric lines at very low R (Bus 2010). The instrument design study needs to understand the influence that R has on the ability to calibrate both wavelength and source flux, including the ability to remove telluric lines and other stray sources. To ensure that the maximum number of objects can be studied while retaining sufficient data return for taxonomic classification, the authors recommend a spectral R of approximately 30 across the CCD response band as a starting point for the design.

One characteristic of single prisms is that the actual R will vary as a function of wavelength, with a higher R in the blue (perhaps by as much as a factor of 4) than in the red. This makes data reduction difficult, while simultaneously adding another factor that will vary the limiting magnitude with wavelength. This can be addressed using multiple prisms to correct the R with wavelength, producing a relatively constant R across the entire CCD response band. The number of prisms must be kept small, both to keep the cost, size, and weight of the instrument reasonable, and to maintain high optical throughput. However, using modern glasses and coatings, this should be possible with three prisms.

Part of the design process should be to simulate the instrument's performance on the telescope before the instrument is built. As an example, we assume sustained access to a 1.5-m telescope to enable characterization studies in a continuing research program. In such a program, one would want to obtain at least three spectra of a rapidly rotating body to determine not only the general mineralogical composition of the body, but also whether that composition changes with surface longitude. As a goal, one should be able to achieve an S/N ratio of 30 with this instrument on a 1.5-m telescope using a 20-minute exposure when observing a V ~20 NEO, bearing in mind that a given S/N ratio in the far red (~1 μm) is more difficult to achieve than in the blue.

This goal appears to be reachable using current technology. A low-resolution spectrograph has been studied conceptually (Konidaris 2010). Preliminary estimates suggest that such an instrument employing prisms for R of 100 on a 1.5-m telescope can reach $M_v$ = 19.9 in 20 minutes with an S/N ratio of 30. An instrument designed for R ~30 should reach a fainter limiting magnitude.

The instrument should have calibration standards built in to permit making observations as rapidly and efficiently as possible. These should include line lamps for wavelength calibration, and one or more lamps for overall photometric calibration. Even with built-in standards, proper calibration of asteroid reflectance spectra requires attention to detail (Howell 2010), including dome flats, observations of solar-type stars to remove the solar spectral component, and either observations of the atmospheric lines, or a good atmospheric model. One possibility would be to observe the calibration stars in one or several broadband filters simultaneously with taking the spectrum of the target object, to obtain data useful for spectro-photometric calibration of the spectrum.

The instrument should also have the fewest possible moving parts to lower its initial construction cost, simplify its operation, increase its reliability, and increase its stability. Stability is critical in an instrument designed to be used in a long-term survey (Bus 2010). Simplicity of design and operation aids in achieving the goal of stability. Using prisms as the disperser will eliminate the need for order sorting filters with a grating, so there is probably no need for a filter wheel. If the slit size is chosen properly, there may be no need for a slit wheel either, given the low R, although this requires further study. About the only mechanism might be a movable mirror to insert the calibration lamp(s), though the fluxes from these lamps could be transmitted continuously by fiber, eliminating that mechanism as well.

Some thought should be given to the opto-mechanical design to determine whether all mechanisms could be eliminated, with the possible exceptions of the shutter, an external dust/environmental (daytime) cover, and camera focus, which must be independent of telescope focus (which is focused by the collimator on the slit). As long as all users can agree on the spectral R, there would be no need for a disperser wheel. A similar instrument for a 4-m to 10-m telescope would probably have a disperser wheel with multiple dispersers. While adding the capability to handle a wider variety of programs and to adapt to conditions at the time of the observation, this adds considerable cost and complexity to the instrument. Instruments for such telescopes typically cost $2 million to $30 million, the latter figure far exceeding the cost of a 2.4-m telescope. To keep the cost of our proposed instrument reasonable for a small telescope, we propose a single fixed slit and a single fixed prism disperser, with the disperser composed of multiple prisms to yield a constant R across the CCD detection bandpass.

One subsystem of the instrument that requires considerable thought from the start is how to implement non-sidereal guiding to hold the object on the slit (Bus 2010). NEOs are usually faint targets, for which all available flux is needed to obtain a useful spectrum, and they will typically move an appreciable distance across the sky during an integration. The ephemeris is not well known for some NEOs, so some sort of guiding off the slit jaws should be studied as one of several guiding modes to be implemented. Several issues should be studied for science impact before instrument design begins, including: slit width versus typical PSF FWHM, how to move the object along the slit to obtain sky background measurements, and whether an independent measure of total flux versus time (to obtain the true flux midpoint versus exposure midpoint) is needed.

Despite the drive for simplicity, the instrument should be controlled by software at the observing console in the control room, and should require no manual controls or adjustments at the telescope. Even camera focus should be performed using remote control of a motor. The software should be built in such a way that there is a low-level software layer controlling the instrument. The engineering Graphical User Interface (GUI) can be built on top of this lower layer using a well-defined Applications Programming Interface (API). This same API serves as the interface to the observatory user's GUI that would be developed either at the observatory where the instrument would be used, or by those developing the instrument.

Increasingly, small telescopes are being automated for a variety of astronomical tasks, from remote teaching of undergraduate laboratory courses, to exoplanet survey and follow-up, to NEO follow-up (e.g., Mutel and Fix 2000 and Pepper *et al.* 2007). Although automation of imaging and photometry are far easier than automation of spectroscopy, the latter has been accomplished for the Tennessee State University 2-m telescope at the Fairborn Observatory (Eaton and Williamson 2004). To use a robotic spectroscopic telescope for NEO characterization, one would need a trustworthy ephemeris for the target, an imaging capability in the spectrograph to verify the field pointing at the time the spectrogram is taken, and an internal guide camera, possibly using light reflected from the slit. Robotic operation would appear at this time to be possible only on the brightest targets.

A potential upgrade path, not listed as a requirement above, is to add an NIR channel fed by a visible/IR dichroic. This second channel would nominally cover the band 1–2.5 μm using (most likely) a standard HgCdTe detector and a separate prism disperser designed to yield an appropriately low R across the NIR wavelength range. This would provide additional information for NEO classification as well as the capability to constrain albedos (as discussed above). An important consideration would be whether to extend this capability into the L band (e.g., to 3.5 μm) to permit observing the important water of hydration absorption features around 3 μm (Feierberg *et al.* 1985). It would require further study to determine whether an R of 30 in this instrument would yield useful information in this wavelength region, but the implementation of such an extension to the basic concept does not appear to pose any serious technical obstacles. To extend the detected wavelengths beyond 2.5 μm, one could use either a 5-micron HgCdTe detector (still somewhat experimental) or an InSb detector. Both would require additional cooling beyond what the 2.5 μm HgCdTe detector would require, so this requirement (and additional cost) would need to be traded against the additional science gain, if any.

Despite the need for simplicity, thought should be given to adding direct imaging capability. This would be used to obtain precise photometric data on NEOs for spectro-photometric calibration, as well as to improve the photometry of NEOs on those nights with sufficient transparency. This is in response to the need for more accurate NEO H magnitudes than typically reported by the NEO discovery surveys and by follow-up astrometrists (Binzel 2010). Such a capability would preclude using graded coatings on the CCD detector with wavelength response that varies with row number, or would require a separate imaging detector that would increase the cost of the instrument. These relatively new graded CCD AR coatings tend to enhance the QE curve of the detector across the entire CCD response band by a few percent over more conventional coatings, but since they vary in response according to the pixel location on the detector, they cannot be used in photometric imaging applications.

Finally, the instrument would be useless without a data reduction pipeline (DRP). An all too common error among instrumentation projects is to exclude the DRP from initial project planning and to assume this critical software can be written in a few weeks by a graduate student working part-time. Instead, the requirements for the DRP should be developed as part of the overall instrument science and engineering requirements, and the DRP should be developed in parallel with the instrument as part of the overall effort, so it is ready and fully functional when the instrument reaches the telescope. Part of this effort is to develop simulated data for the DRP to process during early testing stages, followed by testing the DRP using data taken from the instrument during the laboratory integration and test (I&T) phase. This will aid those trying to understand problems encountered during I&T, as well as speed the commissioning phase. The DRP will also serve observers well, both those at the telescope as well as ensuring that data gathered robotically as part of a large survey project are reduced in a timely fashion. It is expected as part of the DRP development that the DRP will evolve as more is learned about instrument performance and behavior.

## 4. Summary

The science case for a prism-based R ~30 visible-band spectrograph has been presented. From this science case, the basic requirements and concept for an NEO characterization instrument have been derived. Further study is needed to obtain a set of specific science and design requirements based on the science goals for the instrument and the capabilities provided by current technology. Such an instrument routinely used on small telescopes has the potential to vastly increase the quantity and quality of data on NEOs and, in turn, increase the total number of NEOs with taxonomic classifications.

## 5. Acknowledgements

The authors wish to thank S. J. Bus for his informal review of this paper and several constructive comments, R. Binzel and E. Howell for their useful comments, and N. Konidaris for his comments and ideas during the paper's development.